\documentclass{article}
\usepackage{verbatim,amsmath,eufrak,cite}
\usepackage{hyperref}

\hypersetup{
    bookmarks=true,         
    unicode=false,          
    pdftoolbar=true,        
    pdfmenubar=true,        
    pdffitwindow=false,     
    pdfstartview={FitH},    
    pdftitle={My title},    
    pdfauthor={Author},     
    pdfsubject={Subject},   
    pdfcreator={Creator},   
    pdfproducer={Producer}, 
    pdfkeywords={keywords}, 
    pdfnewwindow=true,      
    colorlinks=false,       
    linkcolor=red,          
    citecolor=green,        
    filecolor=magenta,      
    urlcolor=cyan           
}
\def\hhref#1{\href{http://arxiv.org/abs/#1}{arXiv:#1}} 

\renewcommand{\arraystretch}{1.3}
\makeatletter
\newdimen\normalarrayskip              
\newdimen\minarrayskip                 
\normalarrayskip\baselineskip
\minarrayskip\jot
\newif\ifold             \oldtrue            \def\new{\oldfalse}
\def\arraymode{\ifold\relax\else\displaystyle\fi} 
\def\eqnumphantom{\phantom{(\theequation)}}     
\def\@arrayskip{\ifold\baselineskip\z@\lineskip\z@
     \else
     \baselineskip\minarrayskip\lineskip2\minarrayskip\fi}
\def\@arrayclassz{\ifcase \@lastchclass \@acolampacol \or
\@ampacol \or \or \or \@addamp \or
   \@acolampacol \or \@firstampfalse \@acol \fi
\edef\@preamble{\@preamble
  \ifcase \@chnum
     \hfil$\relax\arraymode\@sharp$\hfil
     \or $\relax\arraymode\@sharp$\hfil
     \or \hfil$\relax\arraymode\@sharp$\fi}}
\def\@array[#1]#2{\setbox\@arstrutbox=\hbox{\vrule
     height\arraystretch \ht\strutbox
     depth\arraystretch \dp\strutbox
     width\z@}\@mkpream{#2}\edef\@preamble{\halign
\noexpand\@halignto
\bgroup \tabskip\z@ \@arstrut \@preamble \tabskip\z@ \cr}%
\let\@startpbox\@@startpbox \let\@endpbox\@@endpbox
  \if #1t\vtop \else \if#1b\vbox \else \vcenter \fi\fi
  \bgroup \let\par\relax
  \let\@sharp##\let\protect\relax
  \@arrayskip\@preamble}
\def\eqnarray{\stepcounter{equation}%
              \let\@currentlabel=\theequation
              \global\@eqnswtrue
              \global\@eqcnt\z@
              \tabskip\@centering
              \let\\=\@eqncr
 \halign to \displaywidth\bgroup
    \eqnumphantom\@eqnsel\hskip\@centering
    $\displaystyle \tabskip\z@ {##}$%
    \global\@eqcnt\@ne \hskip 2\arraycolsep
         $\displaystyle\arraymode{##}$\hfil
    \global\@eqcnt\tw@ \hskip 2\arraycolsep
         $\displaystyle\tabskip\z@{##}$\hfil
         \tabskip\@centering
    &{##}\tabskip\z@\cr}
\begingroup\ifx\undefined\newsymbol \else\def\input#1 {\endgroup}\fi

\newcounter{app}

\def\app{\setcounter{equation}{0}
\def\theequation{A\Roman{app}.\arabic{equation}}\par
   \addvspace{4ex}
   \@afterindentfalse
  \secdef\@app\@dapp}
\newcommand\@app{\@startsection {app}{1}{0ex}%
                                   {-3.5ex \@plus -1ex \@minus -.2ex}%
                                   {2.3ex \@plus.2ex}%
                                   {\normalfont\Large\bf}}

\def\@dapp#1{%
{\parindent \z@ \raggedright  \bf #1}\par\nobreak}
\def\l@app#1#2{\ifnum \c@tocdepth >\z@
    \addpenalty\@secpenalty
    \addvspace{1.0em \@plus\p@}%
    \setlength\@tempdima{8.5em}%
    \begingroup
      \parindent \z@ \rightskip \@pnumwidth
      \parfillskip -\@pnumwidth
      \leavevmode \bfseries
      \advance\leftskip\@tempdima
      \hskip -\leftskip
      #1\nobreak\hfil \nobreak\hb@xt@\@pnumwidth{\hss #2}\par
    \endgroup\fi}
\newcounter{sapp}[app]

\def\sapp{\def\theequation{A\arabic{app}.\arabic{equation}}\par
   \@afterindentfalse
  \secdef\@sapp\@dsapp}
\newcommand\@sapp{\@startsection{sapp}{2}{\z@}%
                                     {-3.25ex\@plus -1ex \@minus -.2ex}%
                                     {1.5ex \@plus .2ex}%
                                     {\normalfont\large\bfseries}}

\def\@dsapp#1{%
{\parindent \z@ \raggedright  \bf #1}\par\nobreak}
\newcommand{\l@sapp}{\@dottedtocline{2}{1.5em}{3em}}

\def\be{\begin{eqnarray}}
\def\ee{\end{eqnarray}}

\def\p{\partial}

\def\beq{\begin{equation}}
\def\eeq{\end{equation}}
\def\ba{\beq\new\begin{array}{c}}
\def\ea{\end{array}\eeq}
\def\be{\ba}
\def\ee{\ea}

\def\Tr{{\rm Tr}\,}

\newfont{\alef}{msbm10 at 11pt}
\newfont {\goth}{eufm10 at 11pt}
\def\mathbb#1{\hbox{{\alef #1}}}

\let\@@savethanks\thanks
\def\thanks#1{\gdef\thefootnote{\alph{footnote}}\@@savethanks{#1}}

\hoffset= - 1.0in         

\bigskip

\title{
\bigskip
{\bf
Cut-and-Join Operator Representation for Kontsevich-Witten tau-function} \vspace{.5cm}}
\author{{\bf A. Alexandrov}\thanks{E-mail:  {\tt alexandrovsash at gmail.com}}
\date{ } \\ {\small
{\it CEA, IPhT, 91191 Gif-sur-Yvette, France \&}}\\
 {\small
{\it Ecole Normale Superieure, LPT, 75231 Paris , France \&
}}\\
 {\small
{\it ITEP, Moscow, Russia}}\\
}

\begin{document}

\setcounter{footnote}{0}

\setcounter{tocdepth}{3}

\maketitle

\vspace{-8.0cm}

\begin{center}
\hfill ITEP/TH-35/10\\
\hfill LPT ENS-10/37\\
\hfill IPHT-t10/146\\
\end{center}

\vspace{5.5cm}
\bigskip

\begin{abstract}
In this short note we construct a simple cut-and-join operator representation for Kontsevich-Witten tau-function that is the partition function of the two-dimensional topological gravity. Our derivation is based on the Virasoro constraints. Possible applications of the obtained representation are discussed.
\end{abstract}

\bigskip

\bigskip

\bigskip


\def\thefootnote{\arabic{footnote}}

\section*{Introduction}
\def\theequation{\arabic{equation}}
\setcounter{equation}{0}

Since the early nineties, when the spectacular Witten's conjecture \cite{Witten} was proved by Kontsevich \cite{Konts}, Kontsevich-Witten tau-function, that is the partition function of two-dimensional topological gravity, became an inevitable part of mathematical physics. It arguably can be considered as an elementary building block for more complicated partition functions of the ``Generalized Topological String Theory,'' which unifies conventional topological strings with other theories of the topological/combinatorial invariants possessing universal integrable properties.\footnote{However see \cite{OP}, where it is claimed that the partition function of $CP^1$ model should be considered as the most elementary one and Kontsevich-Witten tau-function is just its particular limit.}
As a consequence of its special role the Kontsevich-Witten partition function is very well studied and a lot of elements of universal description, which are still lacking for more complicated models, are known in this case. Among such interrelated elements are Virasoro constraints, integrable properties, moment variables description, random partitions representation, spectral-curve-based description, a vast net of connections with other models and, of course, Kontsevich matrix integral representation.
This famous matrix integral
is probably the most convenient explicit expression for the Kontsevich-Witten tau-function. The matrix integral representation is very useful in some cases, in particular for understanding of the integrable properties or derivation of the Virasoro constraints. In other cases such representation is not so efficient. The reason is that Kontsevich matrix integral explicitly depends on the external matrix but not on the KdV times, which are Miwa combinations of matrix eigenvalues.

In this paper we derive a simple representation of the Kontsevich-Witten tau-function in terms of the time variables. We claim that the existence of the representation of this type is one more universal property of the topological/combinatorial partition functions. Obtained representation resembles the cut-and-joint operator representations for the generating functions of the single Hurwitz numbers \cite{HHK,MMHodge} and for Hermitian matrix model \cite{Morsh}.
Namely, the tau-function is given by the action of the exponent of the second order differential operator on the trivial initial conditions:
\begin{equation}
\addtolength{\fboxsep}{5pt}
\boxed{
\begin{gathered}
\ \ \ Z_K=e^{\hat W} \cdot 1 \ \ \
\end{gathered}
}
\end{equation}
where the operator is
\be
\hat W= \frac{2}{3}\sum_{k=1}^\infty \left(k+\frac{1}{2}\right)\tau_k\hat L_{k-1}=\\
=\frac{2}{3}\sum_{\substack{k,m\geq 0\\
k+m>0}}\left(k+\frac{1}{2}\right)\left(m+\frac{1}{2}\right)\tau_k\tau_m\frac{\p}{\p \tau_{k+m-1}}+\frac{g^2}{12}\sum_{k,m\geq 0}\left(k+m+\frac{5}{2}\right)
\tau_{k+m+2}\frac{\p^2}{\p\tau_k\p \tau_m}+\frac{1}{g^2}\frac{\tau_0^3}{3!}+\frac{\tau_1}{16}
\ee

Our derivation is based on the existence of Virasoro constraints and homogeneity properties of the partition function. The derivation is very straightforward and directly follows the derivation by Morozov and Shakirov \cite{Morsh} of the similar representation for another important matrix model, namely for the Hermitean one.

\section{The formula}
Kontsevich matrix integral\footnote{For review and the list of the references see e.g. \cite{IZ,ammp,Morint}.}
\be
Z_K=\frac{\int \left[d \Phi\right]\exp\left({-\frac{1}{g}\Tr\left(\frac{\Phi^3}{3!}+\frac{\Lambda \Phi^2}{2}\right)}\right)}{\int \left[d \Phi\right]\exp\left({-\frac{1}{g}\Tr\frac{\Lambda \Phi^2}{2}}\right)}
\label{matint}
\ee
gives a particular representation of the Kontsevich-Witten tau-function with times given by Miwa variables
\be
\tau_k=\frac{g}{(2k+1)}\Tr\frac{1}{\Lambda^{2k+1}}
\ee
It is assumed that the tau-function depends on the infinite set of the independent time variables $\tau$, that mins that the size of the matrix tends to infinity.
In this limit the integral satisfies to an infinite set of the differential equations known as the Virasoro constraints
\be
\hat{L}_n Z_K=\frac{\p}{\p \tau_{n+1}}Z_K,~~~n\geq-1
\label{vir}
\ee
where the operators
\be
\hat{L}_m=\sum_{k=1}^\infty \left(k+\frac{1}{2}\right)\tau_k\frac{\p}{\p \tau_{k+m}}+\frac{g^2}{8}\sum_{k=0}^{m-1}\frac{\p^2}{\p \tau_k \p \tau_{m-k-1}}+\frac{\tau_0^2}{2g^2}\delta_{m,-1}+\frac{1}{16}\delta_{m,0},~~~m\geq -1
\ee
constitute a subalgebra of the Virasoro algebra:
\be
\left[\hat{L}_n,\hat{L}_m\right]=(n-m)\hat{L}_{n+m}
\ee
Let us introduce a graduation, so that $\deg \tau_k =\frac{2k+1}{3}$
and $\deg g=0$. Then the partition function, as a formal series in $\tau$, consists of terms with non-negative degree
\be
Z_K=\sum_{k=0}^\infty Z_K^{(k)}
\ee
with $\deg Z_K^{(k)}=k$. Each $Z_K^{(k)}$ is a polynomial in variables $\tau$, in particular it is obvious that $Z_K^{(0)}=1$. From the point of view of the matrix integral (\ref{matint}) this expansion is very natural -- it is an expansion of the potential, namely:
\be
Z^{(k)}_K=\frac{(-1)^k}{k!(3!g)^k}\frac{\int \left[d \Phi\right]\left(\Tr \Phi^3\right)^k\exp\left(-\frac{1}{g}\Tr\frac{\Lambda \Phi^2}{2}\right)}{\int \left[d \Phi\right]\exp\left({-\frac{1}{g}\Tr\frac{\Lambda \Phi^2}{2}}\right)}
\ee
Let us introduce the degree operator
\be
\hat{D}=\frac{2}{3}\sum_{k=0}^\infty\left(k+\frac{1}{2}\right)\tau_k\frac{\p}{\p \tau_k}
\ee
so that
\be
\hat{D}Z_K^{(k)}=k Z_K^{(k)}
\ee
At the same time from the Virasoro constraints (\ref{vir}) it follows that
\be
\hat{W}Z_K=\hat{D} Z_K
\label{dasw}
\ee
where
\be
\hat{W}=\frac{2}{3}\sum_{k=1}^\infty \left(k+\frac{1}{2}\right)\tau_k\hat L_{k-1}=\\
=\frac{2}{3}\sum_{\substack{k,m\geq 0\\
k+m>0}}\left(k+\frac{1}{2}\right)\left(m+\frac{1}{2}\right)\tau_k\tau_m\frac{\p}{\p \tau_{k+m-1}}+\frac{g^2}{12}\sum_{k,m\geq 0}\left(k+m+\frac{5}{2}\right)
\tau_{k+m+2}\frac{\p^2}{\p\tau_k\p \tau_m}+\frac{1}{g^2}\frac{\tau_0^3}{3!}+\frac{\tau_1}{16}
\label{MastOp}
\ee
Since $\deg \hat{L}_k=-\frac{2m}{3}$ the degree of the operator $\hat{W}$ is equal to one:
\be
\deg \hat{W}=1
\ee
Now taking the terms of the same degree from the left hand side and the right hand side of (\ref{dasw}) one gets
\be
\hat{W} Z_K^{(k)}=(k+1)Z_K^{(k+1)}
\ee
from where it follows that
\be
Z_K^{(k)}=\frac{\hat{W}^k}{k!}Z_K^{(0)}=\frac{\hat{W}^k}{k!}\cdot 1
\ee
After summation over $k$ we obtain the desired formula
\begin{equation}
\addtolength{\fboxsep}{5pt}
\boxed{
\begin{gathered}
\ \ \ Z_K=e^{\hat{W}} \cdot 1 \ \ \
\end{gathered}
}
\label{Mastfor}
\end{equation}
Applying the obtained expression one can easily get an explicit form of the low degree terms of the tau-function, for example:
\be
Z_K^{(1)}=\hat{W}\cdot 1 =\frac{1}{g^2}\frac{\tau_0^3}{3!}+\frac{\tau_1}{16}\\
Z_K^{(2)}=\frac{\hat{W}^2}{2}\cdot 1 ={\frac {1}{72}}\,{
\frac {{\tau_{{0}}}^{6}}{{g}^{4}}}+{
\frac {25}{96}}\,{\frac {{\tau_{{0}}}^{3}\tau_{{1}}}{g^2}}
+{\frac {25}{512}}\,{\tau_{{1}}}^{2}+{\frac {5}{32}}\,\tau_{{0}}\tau_{{2}}
\ee
Of course, these expressions coincide with the known (see e.g. \cite{IMMM,ammp}) low degree terms of the tau-function.\footnote{Using MAPLE we managed to compute terms up to degree 6, and the cut-and-join operator representation turnend out to be rather efficient for the explicit calculations.}

\section{Conclusion and open questions}
In this short note we have derived a simple operator representation for the Kontsevich-Witten tau-function. We claim that obtained representation might be useful for subsequent development of the theory of topological and combinatorial invariants, in particular for understanding of the decomposition formulas \cite{ammp,Giv,Hdec,KosOr,IMMM}. As Kontsevich-Witten tau-function is a key element of the decomposition formulas, the cut-and-join type representation for it allows one to consider on equal footing both intertwining operators and partition functions. The only difference between them is that the intertwining operators are quadratic in the current components and correspond to the quantization of the quadratic hamiltonians \cite{Giv}, while partition functions are given by an action of the cut-and-join operators, which are cubic or of higher power in the current components and perhaps correspond to the quantization of the cubic or higher power hamiltonians.

For instance, now the decomposition formula for the Gaussian branch of the Hermitean matrix model \cite{Hdec,IMMM} can be reformulated purely in terms of the cut-and-join-type operators action on the trivial initial conditions.
The same is true for another relation of similar nature, which connects the Kontsevich-Witten tau-function and the generating function of simple Hurwitz numbers via Hodge integrals \cite{Giv,HHK,MMHodge,MarBouch}. In particular, representation of the form (\ref{Mastfor}) for the generating function of Hodge numbers follows immediately from the known connection with the Kontsevich-Witten tau-function. The only modification is the change of the cut-and-joint operator, which for the Hurwitz numbers is given by
\be
\hat{W}_H=\hat{U}\hat{W}\hat{U}^{-1}
\ee
where $\hat{U}$ is as in \cite{MMHodge}. To compare this representation with the standard cut-and-join operator of the single Hurwitz numbers one should be able to deal with not invertible change of variables \cite{HHK,MMHodge}. This obstacle probably indicates an existence of the natural deformation by some (infinite) set of times.

One more way to use the decomposition formula is to represent  the partition function in
terms of the moment variables.
For the Kontsevich-Witten tau-function this representation is well-known \cite{IZ}. Unfortunately, the representation obtained in this paper does not manifest polynomial structure of the higher genus free energy in terms of the moment variables. Our attempts to modify operator formula so that polynomial structure becomes transparent were not successful.

Is spite of the transparency of our derivation there are still some open questions, both conceptual and technical. Let us mention several properties, which are important for the usual  cut-and-join operator of \cite{HHK,MMHodge} and not obvious for (\ref{MastOp}).
First of all, it is not quite clear to us, if the obtained operator has any natural meaning in the corresponding to KdV hierarchy $\widehat{sl_2}$ subalgebra of $\widehat{gl(\infty)}$. To the best of our understanding, operator (\ref{MastOp}) does not belong to this symmetry subalgebra of the universal Grassmanian. Second, cut-and-join operator of the Hurwitz theory acts in the natural way on the space of the symmetric function, but for the operator (\ref{MastOp}) such action is not known at the moment.

It would be useful to have similar cut-and-join representations for other matrix models, such as complex matrix model, BGW model and Generalized Kontsevich Models. It might be even more interesting to construct the cut-and-join representations for different partition functions of topological strings, for which Kontsevich-Witten tau-function is the simplest example. The first natural candidate here would be $CP^1$. Since in this case Virasoro constraints do not fix the partition function unambiguously, it is hardly possible to derive cut-and-join representation from them. On the other hand Virasoro constraints are powerful enough to restore a full partition function from the function dependent only on the stationary sector variables \cite{Nik}, which is much more simple and is naturally given by the random partition model. We claim that this reconstruction can be done by a particular operator of the cut-and-joint type constructed from the Virasoro constraints.

 It would be also interesting to clarify the connection of the obtained  representation with another approaches to the subject, especially with Eynard technique \cite{Eyn}. We are going to develop some of the above mentioned topics in the subsequent publications \cite{ammta}.


\section*{Acknowledgments}
We are indebted to Andrei Mironov, Alexei Morozov and Dima Panov for useful discussions.
Our work is partly supported by ANR project GranMa "Grandes Matrices Al\'{e}atoires" ANR-08-BLAN-0311-01,
by RFBR grants 09-02-93105-CNRSL and 10-02-00509, grant MK-2567.2011.2 and by Ministry of Education and Science of the Russian Federation
under contract 14.740.11.0081.


\begin{thebibliography}{12}
\bibitem{Witten}
  E.~Witten,
  Surveys Diff.\ Geom.\  {\bf 1} (1991) 243.

\bibitem{Konts}
  M.~Kontsevich,
  Commun.\ Math.\ Phys.\  {\bf 147} (1992) 1.

\bibitem{OP}
  A.~Okounkov and R.~Pandharipande,
  ``The equivariant Gromov-Witten theory of $P^1$,''
  \hhref{math/0207233}.

\bibitem{HHK}
I.~P.~Goulden and D.~M.~Jackson,
Proc. A.M.S.,
{\bf125} (1997), 51--60; \\
R.~Vakil,
 ``Enumerative geometry of curves via degeneration methods,''
Harvard Ph.D. thesis, 1997;  \\
I.~Goulden, D.~Jackson, R.~Vakil,
``The Gromov-Witten potential of a point, Hurwitz numbers, and Hodge integrals,''
 \hhref{math/9910004};\\
M.~Kazarian.
``KP hierarchy for Hodge integrals,''
\hhref{0809.3263}[math];\\
  A.~Mironov, A.~Morozov and S.~Natanzon,
  ``Complete Set of Cut-and-Join Operators in Hurwitz-Kontsevich Theory,''
  \hhref{0904.4227}[hep-th];\\
  B.~Eynard, M.~Mulase and B.~Safnuk,
  ``The Laplace transform of the cut-and-join equation and the Bouchard-Marino
  conjecture on Hurwitz numbers,''
  \hhref{0907.5224}[math.AG];\\
  T.~W.~Brown,
  JHEP {\bf 1005}, 058 (2010)
  \hhref{1002.2099}[hep-th].

\bibitem{MMHodge}
A.~Mironov and A.~Morozov,
  JHEP {\bf 0902}, 024 (2009)
  \hhref{0807.2843}[hep-th].

\bibitem{Morsh}
  A.~Morozov and S.~Shakirov,
  JHEP {\bf 0904} (2009) 064
  \hhref{0902.2627}[hep-th].


\bibitem{IZ}
  C.~Itzykson and J.~B.~Zuber,
  Int.\ J.\ Mod.\ Phys.\  A {\bf 7} (1992) 5661
  \hhref{hep-th/9201001}.

\bibitem{Morint}
A.~Morozov,
  Phys.\ Usp.\  {\bf 37} (1994) 1
  \hhref{hep-th/9303139}.

\bibitem{ammp}
 A.~S.~Alexandrov, A.~Mironov, A.~Morozov and P.~Putrov,
  Int.\ J.\ Mod.\ Phys.\  A {\bf 24} (2009) 4939
  \hhref{0811.2825}[hep-th].

\bibitem{IMMM}
   A.~S.~Alexandrov, A.~Mironov and A.~Morozov,
  Physica D {\bf 235} (2007) 126
 \hhref{hep-th/0608228}.



\bibitem{Giv}
A.~Givental, Int. Math. Res. Not. 2001, No. 23, 1265-1286 (2001),
\hhref{math/0008067};
 Mosc. Math. J. 1, No. 4, 551-568 (2001),
\hhref{math/0108100}.

\bibitem{Hdec}
L.~Chekhov,
  ``Matrix Models and Geometry of Moduli Spaces,''
 \hhref{hep-th/9509001};\\
A.~S.~Alexandrov, A.~Mironov and A.~Morozov,
  Teor.\ Mat.\ Fiz.\  {\bf 150} (2007) 179
  \hhref{hep-th/0605171};\\
  I.~Kostov,
  Nucl.\ Phys.\  B {\bf 837} (2010) 221
  \hhref{0912.2137}[hep-th].




\bibitem{KosOr}
  I.~Kostov and N.~Orantin,
  ``CFT and topological recursion,''
   \hhref{1006.2028}[hep-th].


\bibitem{MarBouch}
V.~Bouchard and M.~ Marino,
 ``Hurwitz numbers, matrix models and enumerative geometry,''
In: From Hodge Theory to Integrability and tQFT: tt*-geometry,
  Proceedings of Symposia in Pure Mathematics, AMS (2008),
\hhref{0709.1458}[math.AG].


\bibitem{Nik}
  N.~A.~Nekrasov,
  Lett.\ Math.\ Phys.\  {\bf 88} (2009) 207.

\bibitem{Eyn}
  B.~Eynard,
  JHEP {\bf 0411} (2004) 031
  \hhref{hep-th/0407261};\\
  B.~Eynard and N.~Orantin,
  ``Invariants of algebraic curves and topological expansion,''
  \hhref{math-ph/0702045};\\
  L.~Chekhov and B.~Eynard,
  JHEP {\bf 0603} (2006) 014
  \hhref{hep-th/0504116}.



\bibitem{ammta}
A.~Alexandrov, A.~Mironov and A.~Morozov,
Cut-and-join operators, matrix models, and characters, to appear.


\end{thebibliography}
\end{document}